\documentclass[%
aps,
prx,
twocolumn,
reprint,
english,
superscriptaddress,
floatfix,
longbibliography,
amsmath,
amssymb]{revtex4-2}

\usepackage{graphicx}
\makeatletter
\DeclareRobustCommand*\uell{\mathpalette\@uell\relax}
\newcommand*\@uell[2]{
  \setbox0=\hbox{$#1\ell$}
  \setbox1=\hbox{\rotatebox{10}{$#1\ell$}}
  \dimen0=\wd0 \advance\dimen0 by -\wd1 \divide\dimen0 by 2
  \mathord{\lower 0.1ex \hbox{\kern\dimen0\unhbox1\kern\dimen0}}
}
\DeclareRobustCommand*\uu{\mathpalette\@uu\relax}
\newcommand*\@uu[2]{
  \setbox0=\hbox{$#1\textit{u}$}
  \setbox1=\hbox{\rotatebox{6}{$#1\textit{u}$}}
  \dimen0=\wd0 \advance\dimen0 by -\wd1 \divide\dimen0 by 2
  \mathord{\lower 0.1ex \hbox{\kern\dimen0\unhbox1\kern\dimen0}}
}

\usepackage{array}
\newcolumntype{M}[1]{>{\centering\arraybackslash}m{#1}}
\newcolumntype{N}{@{}m{0pt}@{}}

\usepackage[utf8]{inputenc} 
\usepackage[T1]{fontenc}
\usepackage[looser]{newtxtext}
\usepackage{microtype}
\usepackage[defaultlines=2,all]{nowidow}
\setcitestyle{numbers,square,citesep={,\kern-.20em}}

\usepackage[normalem]{ulem}
\newcommand\redsout{\bgroup\markoverwith{\textcolor{red}{\rule[0.5ex]{2pt}{1pt}}}\ULon}
\newcommand{\stkout}[1]{\ifmmode\text{\redsout{\ensuremath{#1}}}\else\redsout{#1}\fi}

\usepackage{dcolumn}
\usepackage[dvipsnames]{xcolor}
\usepackage[colorlinks,urlcolor=Blue,linkcolor=Blue,citecolor=Blue]{hyperref}
\usepackage{gensymb}
\usepackage{textcomp}
\usepackage{braket}
\usepackage[nameinlink,capitalise]{cleveref}
\crefname{equation}{Eq.\!}{Eqs.\!}

\newcommand*{\figref}[2][]{%
  \hyperref[{#2}]{%
    Fig.\,\ref*{#2}%
    \ifx\\#1\\%
    \else
      (#1)%
    \fi
  }%
}

\newcommand*{\figureref}[2][]{%
  \hyperref[{#2}]{%
    Figure~\ref*{#2}%
    \ifx\\#1\\%
    \else
      (#1)%
    \fi
  }%
}

\usepackage{bm}
\usepackage{mathtools}
\DeclarePairedDelimiter{\abs}{\lvert}{\rvert}
\DeclarePairedDelimiterX\braopket[3]{\langle}{\rangle}{#1\,\delimsize\vert\,\mathopen{}#2\,\delimsize\vert\,\mathopen{}#3}
\DeclarePairedDelimiter\expect{\langle}{\rangle}
\DeclarePairedDelimiter\floor{\lfloor}{\rfloor}

\DeclarePairedDelimiterXPP\order[1]{\mathcal{O}}{\lparen}{\rparen}{}{#1}

\newcommand{\SrClock}{\ensuremath{{^1\mathrm{S}_0}\!-\!{^3\mathrm{P}_0}}~}%

\usepackage{CJK}
\usepackage{orcidlink}

\begin{document}

\title{Cumulative Fidelity of LMT Clock Atom Interferometers in the Presence of Laser Noise}

\begin{CJK*}{UTF8}{gbsn}

\author{Yijun Jiang (姜一君)\,\orcidlink{0000-0002-1170-7736}}
\affiliation{Department of Physics, Stanford University, Stanford, California 94305, USA}

\author{Jan Rudolph\,\orcidlink{0000-0002-9083-162X}}
\affiliation{Department of Physics, Stanford University, Stanford, California 94305, USA}
\affiliation{Fermi National Accelerator Laboratory, Batavia, Illinois 60510, USA}

\author{Jason M.\ Hogan\,\orcidlink{0000-0003-1218-2692}}
\email[]{hogan@stanford.edu}
\affiliation{Department of Physics, Stanford University, Stanford, California 94305, USA}

\begin{abstract}
Clock atom interferometry is an emerging technique in precision measurements that is particularly well suited for sensitivity enhancement through large momentum transfer (LMT). While current systems have demonstrated momentum separations of several hundreds of photon momenta, next-generation quantum sensors are targeting an LMT enhancement factor beyond $10^4$. However, the viability of LMT clock interferometers has recently come into question due to the potential impact of laser frequency noise. Here, we resolve this concern by analyzing the cumulative fidelity of sequential state inversions in an LMT atom interferometer. We show that the population error from $n$ pulses applied from alternating directions scales linearly with $n$. This is a significant advantage over the $n^2$ scaling that occurs when probing a two-level system $n$ times from the same direction. We further show that contributions to the interferometer signal from parasitic paths generated by imperfect pulses are negligible, for any loss mechanism. These results establish that laser frequency noise is not a practical limitation for the development of high-fidelity LMT clock atom interferometers.
\end{abstract}

\maketitle
\end{CJK*}

\section{Introduction}
\label{sec:introduction}
Clock atom interferometry is a promising tool for tests of fundamental physics, gravitational wave astronomy in unexplored frequency bands~\cite{yu2011gravitational,graham2013new,graham2016resonant,graham2017mid,tino2019sage,elneaj2020aedge,badurina2020aion,abe2021matter,badurina2022prospective}, searches for ultralight dark matter~\cite{graham2016dark,graham2018spin,arvanitaki2018search,derr2023clock,dipumpo2022light}, tests of the universality of the gravitational redshift~\cite{roura2020gravitational}, tests of the quantum twin paradox~\cite{loriani2019interference}, and quantum superposition at macroscopic scales~\cite{nimmrichter2013macroscopicity,arndt2014testing}. Reaching the required sensitivity for such experiments relies on sensitivity enhancement through large momentum transfer (LMT), which increases the space-time separation between the interferometer arms~\cite{mcguirk2000large,mueller2008atom,mueller2009atom,clade2009large,chiow2011102hk,mcdonald201380hk,kovachy2015quantum,mazzoni2015large,kotru2015large,plotkin2018three,gebbe2019twin,pagel2020symmetric,rudolph2020large,zhou2011development,beguin2023atom,rodzinka2024optimal}. In a clock atom interferometer, this is accomplished using a sequence of pulses from a single laser beam applied from alternating directions. A momentum separation of several hundred photon momenta has been demonstrated on a moderately narrow clock transition~\cite{rudolph2020large,wilkason2022atom}, but further enhancement is limited by spontaneous emission loss due to the short excited state lifetime. Many proposals for future clock atom interferometer implementations~\cite{abe2021matter,badurina2020aion,zhan2020zaiga} target ultranarrow transitions such as the \SrClock line in strontium~\cite{hu2017atom,hu2019sr}, whose extremely long excited state lifetime can in principle support an LMT enhancement beyond $10^4$.

Compared with its two-photon counterpart, the single-photon transitions typically envisioned in clock atom interferometers allow for superior cancellation of laser frequency noise as a common mode in differential readout~\cite{dimopoulos2008general,yu2011gravitational,graham2013new,legouet2007influence}. However, laser frequency noise, along with other noise sources, reduces the transfer efficiency of each pulse~\cite{szigeti2012momentum}, leading to contrast degradation and thus a reduction of phase sensitivity~\cite{hu2017atom,rudolph2020large}. The loss of contrast due to laser frequency noise has previously been understood to scale linearly with the LMT order $n$, such that an rms frequency noise below $10~\text{Hz}$ in the bandwidth of interest is sufficient to reach an LMT order of $10^3\!-\!10^4$~\cite{abe2021matter}. However, a recent paper claims that the loss associated with laser frequency noise in an LMT clock atom interferometer grows with $n^2$~\cite{chiarotti2022practical}, posing much stricter requirements on laser frequency stability and pushing such ambitious LMT targets out of reach of current laser technology.

In this paper, we simulate a narrowband clock atom interferometer where the atoms are interrogated by pulses from alternating directions and the resulting Hilbert space is spanned by both the internal atomic levels and the external momentum states. Each pulse promotes the majority of the wavefunction to a new momentum state and, while the overall population is decreased, the induced population error does not carry over to the next pulse. Summing over errors from $n$ sequential $\pi$ pulses thus leads to a fidelity loss proportional to $n$, for any LMT atom interferometer. In contrast, the previous work implicitly treats the atom in a two-level Hilbert space probed $n$ times from the same direction \cite{chiarotti2022practical}. Then the atom alternates between just two states, leading to an $n^2$ scaling of the population error. This treatment does not apply to an LMT clock atom interferometer, where the Hilbert space is significantly expanded by the external momentum states.

In addition to verifying the linear scaling law for population errors, we explore marginal contributions from the parasitic paths spawned by the pulse errors. We develop a mathematical formalism to group these paths by the number of errors and bound their contrast loss contribution to an insignificant level, regardless of the number of pulses $n$. Additionally, we derive an analytic expression for the frequency noise transfer function of the population loss and calculate the sequence fidelity of $n$ consecutive LMT pulses using a realistic laser noise spectrum. We then put this analysis in the context of a full $n\hbar k$ interferometer and show that, for practical Rabi frequencies, a laser stabilized to an rms frequency noise level below $10~\text{Hz}$ can in principle support an LMT order at the $10^3\!-\!10^4$ level. This conclusion agrees with previous proposals for next-generation long-baseline clock atom interferometers~\cite{abe2021matter} and indicates that laser frequency noise does not pose a practical constraint for LMT clock atom interferometers.

\section{Interferometer fidelity and contrast}
\label{sec:fidelity-contrast}

We consider a narrowband LMT clock atom interferometer, where the applied pulses are generally velocity selective and only address one arm of the interferometer at a time [see \figref[a]{fig:interferometer-diagrams}]. Thus, only the initial and final beamsplitter pulses and the central mirror pulse are common to both arms. In between, augmentation zones of additional $\pi$ pulses are inserted to increase and decrease the momentum separation. This requires consecutive pulses from alternating directions, whose frequencies are continuously adjusted to compensate for the Doppler shift of the atoms. To close an $n\hbar k$ interferometer, a total of $(4n{-}3)$ $\pi$ pulses are required, with each of the four augmentation zones consisting of $n{-}1$ pulses. In the following, we analyze the impact of imperfect consecutive $\pi$ pulses on the fidelity of the interferometer output signal. 

During the first half of the interferometer, the upper arm is accelerated by $\pi$ pulses from alternating directions. Each pulse in this sequence changes the current state by flipping its internal level and increasing its momentum by one photon recoil, while leaving a small fraction of the wavefunction untransferred due to pulse imperfections. Since the pulses are velocity selective, the untransferred part of the wavefunction is Doppler suppressed from interacting with the next pulse, which is frequency tuned to the new momentum state [see \figref[b]{fig:interferometer-diagrams}]. Despite reducing the overall amplitude, this wavefunction error will generally not accumulate momentum for the rest of this augmentation zone and instead drift away from the main interferometer arm path. However, the parasitic paths can interact with future pulses during the deceleration of the interferometer arm and may drift back to interfere with the main paths at the output ports. Such an occurrence would require both spatial overlap with the main output ports and population in the relevant final momentum states (0 or $1\hbar k$).

\begin{figure}[ht]
    \centering
    \includegraphics[width=\linewidth]{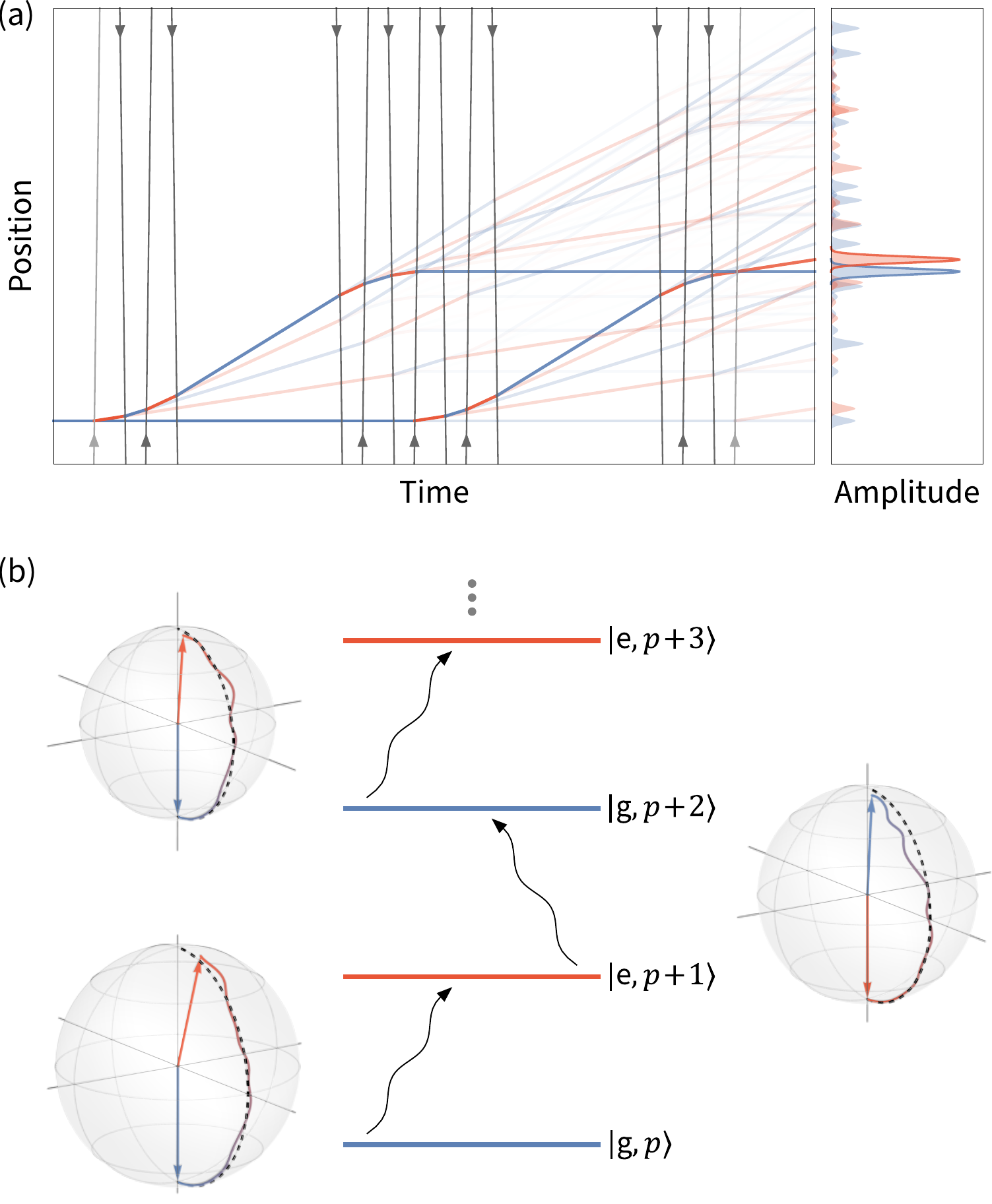}
    \caption{(a) Space-time diagram of a narrowband clock atom interferometer with exaggerated pulse infidelities of $0.05$. The vertical lines represent light pulses traveling in opposite directions indicated by the arrows, with the light travel time exaggerated. The first and last pulses are $\pi/2$ pulses while all others are $\pi$ pulses. The internal levels are color coded in blue (ground) and red (excited), with opacity representing the wavepacket amplitudes of the paths. The position distribution of all path amplitudes after a finite drift time is also shown, with a wavepacket size such that the two output ports are resolved. (b) Bloch sphere representation of an LMT clock atom interferometer with velocity-selective pulses. Each $\pi$ pulse couples two states with opposite internal levels and an external momentum difference of $\hbar k$. The alternation of pulse direction (slanted arrows) builds up momentum in the wavefunction, moving it to a new Bloch sphere after each pulse. Laser frequency noise leads to untransferred population that is not carried over to the next Bloch sphere but merely decreases its size, resulting in a reduction in the radius as shown.}
    \label{fig:interferometer-diagrams}
\end{figure}

Without loss of generality, we can express the population loss in the main paths separately from loss arising from interference due to parasitic paths. Upon interaction, each $\pi$ pulse imprints a phase $\phi_j$, and the pulse imperfection reduces the wavefunction amplitude in a given arm by $\sqrt{1-\delta\!P_j}$, where $\delta\!P_j$ is the population loss from the $j$-th pulse. For a full $n\hbar k$ interferometer, the final state of the upper arm before the closing beamsplitter can be written as
\begin{equation}
    \ket{\Psi^{\uu}}=c^{\uu}\ket{\text{g},0}+\varepsilon_{\text{g},0}^{\uu}\ket{\text{g},0}+\varepsilon_{\text{e},1}^{\uu}\ket{\text{e},1}+\{p\geqslant2\text{ states}\},
    \label{eq:series-parasitic-grouped-upper}
\end{equation}
where the states are labeled by their internal levels $s \in \{\text{g},\text{e}\}$ and external momentum $p\in\mathbb{Z}$ measured in units of $\hbar k$. The superscript $\uu$ denotes the upper arm. The first term is the final state of the main upper path, with amplitude
\begin{equation}
    c^{\uu}\equiv\prod_{j=1}^{2n-1}e^{i\phi_j^{\uu}}\sqrt{1-\delta\!P_j^{\uu}}
    \label{eq:cu-definition}
\end{equation}
and where the subsequent terms describe the parasitic states. Among all possible parasitic paths, we explicitly group those terminating in $\ket{\text{g},0}$ and $\ket{\text{e},1}$, and denote their total amplitudes as $\varepsilon_{\text{g},0}^{\uu}$ and $\varepsilon_{\text{e},1}^{\uu}$. These states will affect the readout after the closing beamsplitter, while all other parasitic states do not have the relevant momentum and are collected in the last term in \Cref{eq:series-parasitic-grouped-upper}. Similarly, the final state of the lower arm before the closing beamsplitter is
\begin{equation}
    \ket{\Psi^{\uell}}=c^{\uell}\ket{\text{e},1}+\varepsilon_{\text{g},0}^{\uell}\ket{\text{g},0}+\varepsilon_{\text{e},1}^{\uell}\ket{\text{e},1}+\{p\geqslant2\text{ states}\},
    \label{eq:series-parasitic-grouped-lower}
\end{equation}
where
\begin{equation}
    c^{\uell}\equiv\prod_{j=2n-1}^{4n-3}e^{i\phi_j^{\uell}}\sqrt{1-\delta\!P_j^{\uell}}
    \label{eq:cl-definition}
\end{equation}
and the superscript $\uell$ denotes the lower arm.

We define the cumulative fidelity of the two arms before the closing beamsplitter as $F^{\uu}\equiv\abs{\braket{\text{g},0|\Psi^{\uu}}}^2$ and $F^{\uell}\equiv\abs{\braket{\text{e},1|\Psi^{\uell}}}^2$. Up to first order in $\varepsilon_{\text{g},0}^{\uu}$ and $\varepsilon_{\text{e},1}^{\uell}$, and assuming $n$ is sufficiently large such that $(2n-1)\,\delta\!P\approx 2n\,\delta\!P$, we obtain
\begin{align}\label{eq:fidelity}
\begin{split}
    F^{\uu}&\approx\left|c^{\uu}\right|^2 + 2\operatorname{Re}\left(\varepsilon_{\text{g},0}^{\uu}\cdot{c^{\uu}}^*\right) \approx 1-2n\,\delta\!P - \delta\!F^{\uu}_{\text{para}}\\
    F^{\uell}&\approx\left|c^{\uell}\right|^2 + 2\operatorname{Re}\left(\varepsilon_{\text{e},1}^{\uell}\cdot{c^{\uell}}^*\right) \approx 1-2n\,\delta\!P - \delta\!F^{\uell}_{\text{para}},
\end{split}    
\end{align}
with average per-pulse population loss $\delta\!P$, and where we define $\delta\!F^{\uu}_{\text{para}}\equiv -2\operatorname{Re}(\varepsilon_{\text{g},0}^{\uu}\cdot{c^{\uu}}^*)$ and $\delta\!F^{\uell}_{\text{para}}\equiv -2\operatorname{Re}(\varepsilon_{\text{e},1}^{\uell}\cdot{c^{\uell}}^*)$ as the contributions of parasitic paths that may interfere with the desired output ports of the interferometer. After the final beamsplitter with phase angle $\theta$, the contrast of an $n\hbar k$ interferometer $C$ emerges from the interference term $2\operatorname{Re}(e^{i\theta}\Psi^{\uu}{\Psi^{\uell}}^*)$. Since amplitudes in the interference-allowed states scale as the square root of the fidelities, the contrast can be written as the geometric mean of $F^{\uu}$ and $F^{\uell}$. When the parasitic contributions are negligible,
\begin{equation}
    C \approx \sqrt{F^{\uu}\,F^{\uell}} \approx 1-2n\,\delta\!P,\label{eq:contrast}
\end{equation}
which scales linearly with $n$, as expected. This is a manifestation of the extended Hilbert space, which is spanned by not only the internal atomic levels but also the external momentum states, an essential feature of any LMT atom interferometer.

\section{Parasitic interferometer paths}
\label{sec:parasitic-paths}
In this section, we analyze the fidelity loss induced by the parasitic paths and show that it is indeed a small correction for practical LMT clock atom interferometers. As depicted in \figref{fig:interferometer-diagrams}, pulse errors spawn parasitic paths of untransferred atoms, with small amplitudes relative to the main path. We write the final state of a generic parasitic path as $\ket{\psi^{(m)}(j_1,j_2,\cdots,j_m)}$, where $m$ is the number of pulse errors, and $\{j_1,j_2,\cdots,j_m\}$ is the set of indices of all nominally resonant pulses where errors can occur. Taking the upper arm as an example, with superscripts omitted for convenience, \cref{eq:series-parasitic-grouped-upper} can be rewritten by grouping the parasitic terms according to the number of errors:
\begin{align}
    \ket{\Psi}=&\,c\ket{\text{g},0}
    +\sum_{j_1=1}^{4n-3}\alpha_{j_1}\Ket{\psi^{(1)}(j_1)}+\notag\\
    &\sum_{j_1=1}^{4n-3}\sum_{j_2=j_1+1}^{4n-3}\alpha_{j_1}\alpha_{j_2}\Ket{\psi^{(2)}(j_1,j_2)}+\cdots\notag\\
    =&\,c\ket{\text{g},0}+\notag\\
    &\sum_{m\geqslant1}\left(\sum_{j_1=1}^{4n-3}\!\cdots\!\!\!\sum_{\substack{j_m=\\j_{m\!-\!1}+1}}^{4n-3}\!\!\!\alpha_{j_1}\cdots\alpha_{j_m}\Ket{\psi^{(m)}(j_1,\cdots\!,j_m)}\!\right),
    \label{eq:series-parasitic-expanded-upper}
\end{align}
where $\alpha_j$ represents the complex amplitude of the projection error at the $j$-th pulse, which relates to the population loss via $|\alpha_j|^2=\delta\!P_j$. For ease of analysis, we drop the phase and use a universal $\alpha\in\mathbb{R}$ as the error amplitude. This simplification treats all parasitic paths as in phase, and thus leads to an upper bound on their interference at the output ports.

\begin{figure*}[t]
    \centering
    \includegraphics[width=\linewidth]{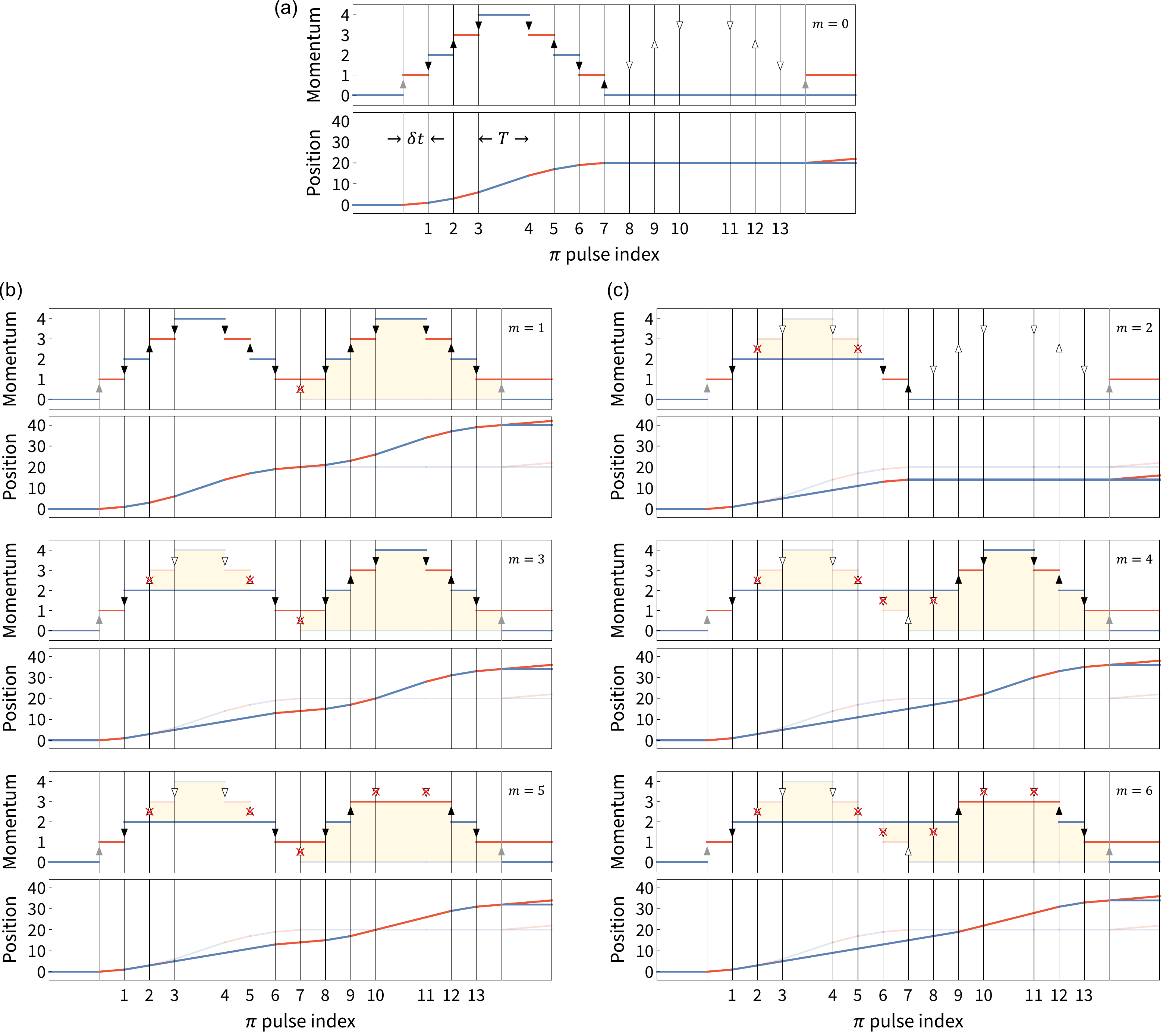}
    \caption{Momentum-time and space-time diagrams of the upper arm of a narrowband $4\hbar k$ clock atom interferometer for different numbers of pulse errors $m$. The internal atomic levels are indicated as blue (ground) and red (excited). The vertical lines represent $\pi$ pulses (black) and $\pi/2$ pulses (gray), with arrows indicating their directions and detunings. We assume equal pulse spacing $\delta t$ and a non-zero interrogation time $T$. The vertical axes are dimensionless, with momentum in units of $\hbar k$ and position in units of $v_r\,\delta t$, where $v_r$ is the recoil velocity. (a) The main upper path without pulse error ($m=0$). Note that the second half of the sequence addresses the lower arm and is off resonant with the main upper path, indicated by the open arrows. (b), (c) Example parasitic paths, where the cross marks indicate pulses with transition errors. Due to these errors, the parasitic paths fail to follow the main path (shown for reference in lower opacity). The yellow-shaded area indicates the relative position error of the parasitic path with respect to the main path, which is also shown explicitly in the space-time diagram. (c) Example paths with even numbers of errors, increasing from top to bottom with $m=2,4,6$. We have selected an $m=4$ example such that the first two errors are the same as the $m=2$ example, and the subsequent errors lead to additional integrated relative position error (indicated by additional yellow shading).  A similar choice is made for the $m=6$ example. (b) Example paths with odd numbers of errors $m=1,3,5$. To satisfy the momentum constraint with an odd $m$, an error must occur at the center mirror pulse. Note that the examples in (b) and (c) are only illustrative, and many other parasitic paths are possible. For a single-loop narrowband Mach-Zehnder interferometer sequence, $m=6$ is the largest possible number of errors, independent of $n$.}
    \label{fig:parasitic-paths}
\end{figure*}

\figureref{fig:parasitic-paths} shows example parasitic paths of a narrowband interferometer with varying number of pulse errors $m$. While the number of parasitic paths grows with $m$, the corresponding amplitude in \cref{eq:series-parasitic-expanded-upper} is suppressed by $\alpha^m$. Counting the number of paths for each $m$ is required to bound its contribution to the series. Due to the linearity of the Schr{\"o}dinger equation, it is possible to treat each path independently and only analyze their interference at the detection region. Note that only parasitic paths that lead to the same final momentum states as the main paths can cause interference, since the other velocity classes have different de Broglie wavelengths. We thus restrict the counting to those $\ket{\psi^{(m)}(j_1,\cdots,j_m)}$ that project onto $\ket{\text{g},0}$ or $\ket{\text{e},1}$, leading to a total number of $N_p(m,n)$ paths, where the subscript indicates the momentum constraint. We also exclude paths with final positions far away from the detection region that do not overlap with the main paths. This further reduces the number of paths to $N_{p,x}(m,n)$, where the subscripts indicate the constraints from both momentum and position.

The final momentum constraint poses a strong bound on the number of relevant paths produced in an LMT sequence (see \cref{appendix:combinatorial}). Pulse errors must occur either at the center mirror pulse [top panel in \figref[b]{fig:parasitic-paths}] or be grouped in pairs with the same detuning [see \figref[c]{fig:parasitic-paths}]. Thus, for a single-loop narrowband Mach-Zehnder interferometer sequence shown in \figref{fig:parasitic-paths}, the series in \cref{eq:series-parasitic-expanded-upper} is truncated at $m=6$. For any specific $m$, only $N_p(m,n)=\order{n^{\floor{m/2}}}$ combinations of pulse error indices are allowed. We validate this scaling numerically by enumerating all possible paths for LMT interferometers up to $n=15$ [see \figref[a]{fig:parasitic-path-contributions}].

The final position constraint further reduces the number of relevant pulse error combinations. We define a detection region with a radius $w$, and only consider paths that terminate within this distance from the main paths. Any paths outside of this region do not cause interference as long as the wavepackets are well localized. Here, we assume $w=1~\text{mm}$, which is a practical experimental setting and is comparable with the wavepacket size of a nanokelvin temperature cloud after an expansion time of a few seconds. The relative position of example parasitic paths with respect to the main upper path is visualized in \figref{fig:parasitic-paths} by the yellow-shaded area in the momentum-time diagram. These figures include a nonzero interrogation time $T$ between the accelerating and decelerating zones for illustrative purposes. However, we also analyze a worst-case bound with zero interrogation time in \cref{appendix:combinatorial}. As shown in \figref[b]{fig:parasitic-paths}, a parasitic path that satisfies the momentum constraint with $m=1$ will never terminate in the vicinity of the main path. For parasitic paths with $2\leqslant m\leqslant6$, the number of possible pulse error combinations is reduced from $N_p(m,n)$ to $N_{p,x}(m,n)=\frac{w}{v_r\,\delta t}\times\order{n^{\floor{m/2}-1}}$, where $v_r$ (typically mm/s) is the recoil velocity and $\delta t$ (typically ms) is the spacing between the rising edges of two adjacent pulses (see \cref{appendix:combinatorial}). To make the pulses nonoverlapping, we require $\delta t\geqslant\pi/\Omega$, where $\Omega$ is the Rabi frequency of the pulses.

For the upper arm, we determine an upper bound of the parasitic-path-induced wavefunction error $\varepsilon$ by adding the amplitudes of all relevant paths in phase, namely those that terminate in $\ket{\text{g},0}$ and $\ket{\text{e},1}$ within the detection region. This is equivalent to summing the number of paths $N_{p,x}(m,n)$ over the number of errors $m$, weighted by an amplitude suppression of $\alpha^m$:
\begin{equation}
    \varepsilon\equiv\left|\varepsilon_{\text{g},0}^{\uu}\right|+\left|\varepsilon_{\text{e},1}^{\uu}\right|=\sum_{m=2}^6\alpha^mN_{p,x}(m,n).
    \label{eq:total-parasitic-wavefunction}
\end{equation}
This expression can be simplified by taking the $m=2$ term in the summation
\begin{equation}
    \varepsilon=\sum_{m=2}^6\alpha^m\frac{w}{v_r\,\delta t}\times\order*{n^{\left\lfloor{m/2}\right\rfloor-1}}\approx\frac{w}{v_r\,\delta t}\times\order*{\alpha^2},
    \label{eq:total-parasitic-wavefunction-bound}
\end{equation}
The $m=2$ term dominates assuming $n\,\alpha^2\ll1$, consistent with the assumption that the loss of contrast is small. However, even in the case where $n\,\alpha^2=\order{1}$, this term is not surpassed by the subsequent terms; thus, \Cref{eq:total-parasitic-wavefunction-bound} still gives the correct bound. Note that \cref{eq:total-parasitic-wavefunction-bound} is equally valid for the lower interferometer arm, so $\varepsilon$ bounds the parasitic-path-induced wavefunction error from the lower arm as well.

To set a bound on the infidelity contribution from the parasitic paths in \Cref{eq:fidelity}, we assume the worst-case scenario where $\varepsilon_{\text{g},0}^{\uu}$ is completely out of phase with $c^{\uu}$, whose magnitude is approximately 1, and likewise for $\varepsilon_{\text{e},1}^{\uell}$ and $c^{\uell}$. Since $\delta\!F^{\uu}_{\text{para}}\leqslant2|\varepsilon_{\text{g},0}^{\uu}|$ and $\delta\!F^{\uell}_{\text{para}}\leqslant2|\varepsilon_{\text{e},1}^{\uell}|$, it is convenient to use a weaker bound that is common for both arms,
\begin{equation}
    \delta\!F_{\text{para}}^a\leqslant2\left(\left|\varepsilon_{\text{g},0}^a\right|+\left|\varepsilon_{\text{e},1}^a\right|\right)=2\varepsilon\,\,\quad a\in\{\uu,\uell\}.
    \label{eq:delta-fidelity-parasitic}
\end{equation}

Note that unlike the population loss in the main path, \cref{eq:total-parasitic-wavefunction-bound,eq:delta-fidelity-parasitic} suggest that the parasitic-path-induced error is bounded by a constant at large LMT order $n$. To explain this, we notice that the suppression from transition error amplitudes $\alpha^m$ is more significant than the growth of $N_{p,x}(m,n)$ with respect to $m$. \Cref{eq:total-parasitic-wavefunction} is thus dominated by $m=2$ paths that satisfy both momentum and position constraints, whose total number $N_{p,x}(2,n)=\frac{w}{v_r\,\delta t}\times\mathcal{O}(1)$ is independent of $n$.

We numerically calculate the parasitic-path-induced amplitude error for LMT interferometers up to $n=15$ in the case of densely packed pulses, such that $\delta t=\pi/\Omega$. \figureref[b]{fig:parasitic-path-contributions} confirms that the error has a constant upper bound for an interferometer operated at Rabi frequency $\Omega=2\pi\times1~\text{kHz}$, recoil velocity $v_r=6.6~\text{mm/s}$, and detection region radius $w=1~\text{mm}$. With pulse inefficiency $\alpha^2=10^{-4}$ and zero interrogation time, the wavefunction error at large $n$ is $\varepsilon=5\times10^{-3}$, which causes a small degradation of the interferometer contrast by at most $2\varepsilon=1\times10^{-2}$. A finite interrogation time $T$ further reduces the loss by posing a stronger position constraint on the parasitic paths.

While in this work we focus on reduction of contrast, parasitic paths also lead to a perturbation of the interferometer phase~\cite{siemss2023large,chen2023enhancing,saywell2025temporal}, which is proportional to the wavefunction error $\varepsilon$ at the output ports. Depending on the nature of the experiment, this phase error could be an important source of systematic error. For the MAGIS-100 experiment, there will be some degree of common-mode suppression of such errors in the gradiometer signal, and phase errors are only significant if they cause fluctuations within the frequency band of interest.

\begin{figure}[t]
    \centering
    \includegraphics[width=\linewidth]{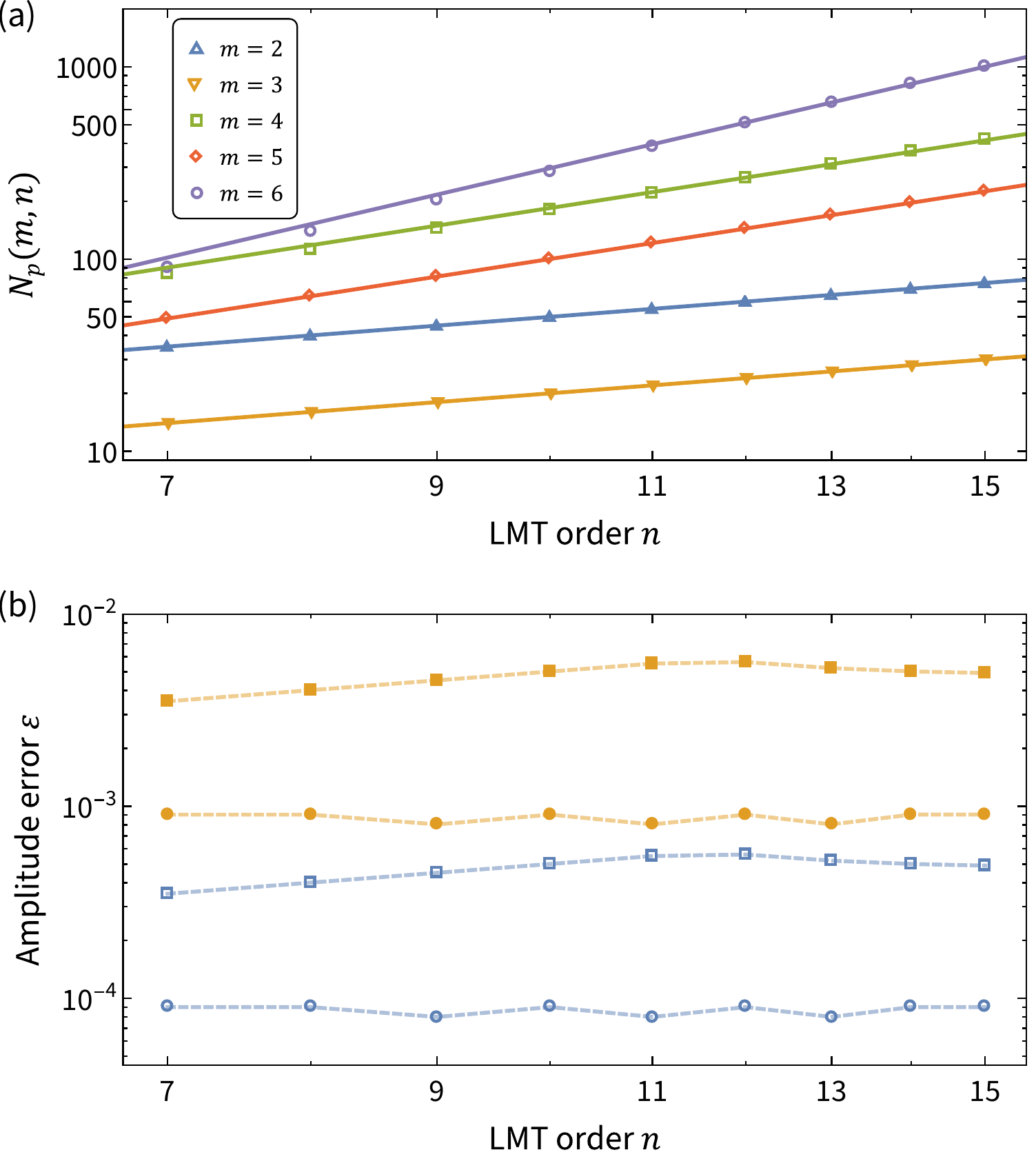}
    \caption{(a) Number of parasitic paths that satisfy the momentum constraint $N_p(m,n)$ as a function of LMT order $n$, obtained by numerically enumerating all paths with a fixed number of pulse errors $m$, and only including those with final momentum $p=0$ or $p=1$. The lines are monomial fits to the power of $\floor{m/2}$, showing good agreement with the predicted $\mathcal{O}(n^{\floor{m/2}})$ growth for sufficiently large $n$ (e.g., $n\geqslant7$). (b) Parasitic-path-induced error $\varepsilon$ as a function of LMT order $n$, obtained by constructively summing over amplitudes of all parasitic paths that satisfy both the momentum constraint and the position constraint. We assume densely packed pulses at $\Omega=2\pi\times1~\text{kHz}$, $w=1~\text{mm}$, and $v_r=6.6~\text{mm/s}$. We show two pulse inefficiency values $\alpha^2=10^{-4}$ (filled orange markers) and $\alpha^2=10^{-5}$ (open blue markers), with interrogation time $T=0~\text{ms}$ (squares) or $T=50~\text{ms}$ (circles).
    The induced error does not grow with $n$ for large enough $n$.
    }
    \label{fig:parasitic-path-contributions}
\end{figure}

\section{Frequency noise transfer function and laser stability requirement}
\label{sec:transfer-function}
Since the interference of the parasitic paths only weakly degrades the sequence fidelity and does not scale with the number of pulses, the dominant effect of laser frequency noise on interferometer contrast is the amplitude loss in the main paths. To study the practical laser stability requirement, we ignore the parasitic paths and derive an analytic transfer function relating the population error in a pulse sequence to the power spectral density of laser frequency noise. As discussed in \cref{appendix:analytic}, for a single square-shaped $\pi$ pulse subject to a weak frequency noise $\delta\nu(t)$, the population error can be expressed as
\begin{equation}
    \delta\!P = \left(\int_{-\infty}^{+\infty} h_1(t)\,\delta\nu(t)\,dt\right)^2,
\end{equation}
where $h_1(t)$ is the second-order time response function. As detailed in \Cref{eq:appendix-b.population-loss-frequency-domain,eq:appendix-b.frequency-response-function,eq:appendix-b.ensemble-average,eq:appendix-b.ensemble-averaged-loss}, we take the ensemble average
\begin{equation}
    \expect{\delta\!P} = \int_{0}^{\infty}\left|\widehat{h_1}(f)\right|^2 S_{\nu}(f)\,df,
\end{equation}
where we Fourier transform $h_1(t)$ and $\delta\nu(t)$ into $\widehat{h_1}(f)$ and $\widehat{\delta\nu}(f)$ and identify
\begin{equation}
    \expect{\widehat{\delta\nu}^*\!(f)\,\widehat{\delta\nu}(f')}=\tilde{S}_{\nu}(f)\,\delta(f-f')=\frac{1}{2}S_{\nu}(f)\,\delta(f-f'),
\end{equation}
where $\tilde{S}_{\nu}(f)$ and $S_{\nu}(f)$ are, respectively, the two-sided and one-sided power spectral density of the frequency noise. We recognize $H_1(f)\equiv|\widehat{h_1}(f)|^2$ as the single-pulse frequency noise transfer function for population loss, whose analytic form for a square pulse is
\begin{equation}
    H_1(f)=\left(\frac{2\pi}{\Omega}\right)^2\left(\frac{\Omega^2\cos\left(\pi f\,\frac{\pi}{\Omega}\right)}{\Omega^2-(2\pi f)^2}\right)^2.
    \label{eq:transfer-function}
\end{equation}

With the interference from parasitic paths ignored, \cref{eq:fidelity} expresses sequence infidelity as a linear accumulation of the population loss from each pulse, which can also be applied to a subsequence of an interferometer arm. Here we focus on a single augmentation zone of the interferometer and study the loss from $n$ consecutive pulses applied from alternating directions. Since the population loss grows linearly with $n$, the $n$-pulse frequency noise transfer function for population loss has the form $H_n(f)=n\,H_1(f)$. Using this transfer function, we calculate the fidelity of the augmentation zone in the presence of laser frequency noise by numerically integrating
\begin{equation}
    F_{\textup{aug}}=1-\int_{f_l}^{\infty}H_n(f)\,S_{\nu}(f)\,df
\end{equation}
for a given laser noise model described below, where $f_l=0.1$\,Hz is the lower cutoff frequency determined by the cycle time of a typical experiment. For numerical feasibility, we approximate the upper cutoff frequency by a finite $f_u$, chosen as the first node of $H_n(f)$ at $3\times\Omega/2\pi$, where the envelope of $H_n(f)$ is reduced to 1.6\% of its dc value. \figureref[b]{fig:transfer-function-and-fidelity} shows the fidelity $F_{\textup{aug}}$ as a function of the rms noise amplitude $\Delta\nu$, with the laser noise power spectral density $S_\nu(f)$ given by the inset and Rabi frequency $\Omega=2\pi\times1~\text{kHz}$. We find that at a moderate laser noise amplitude of $\Delta\nu=10~\text{Hz}$, the per-pulse population loss is $\alpha^2=4.7\times10^{-5}$, leading to a zone fidelity of $63\%$ after $10^4$ pulses, assuming that laser frequency noise is the only source of loss.

\begin{figure}[tb]
    \centering
    \includegraphics[width=\linewidth]{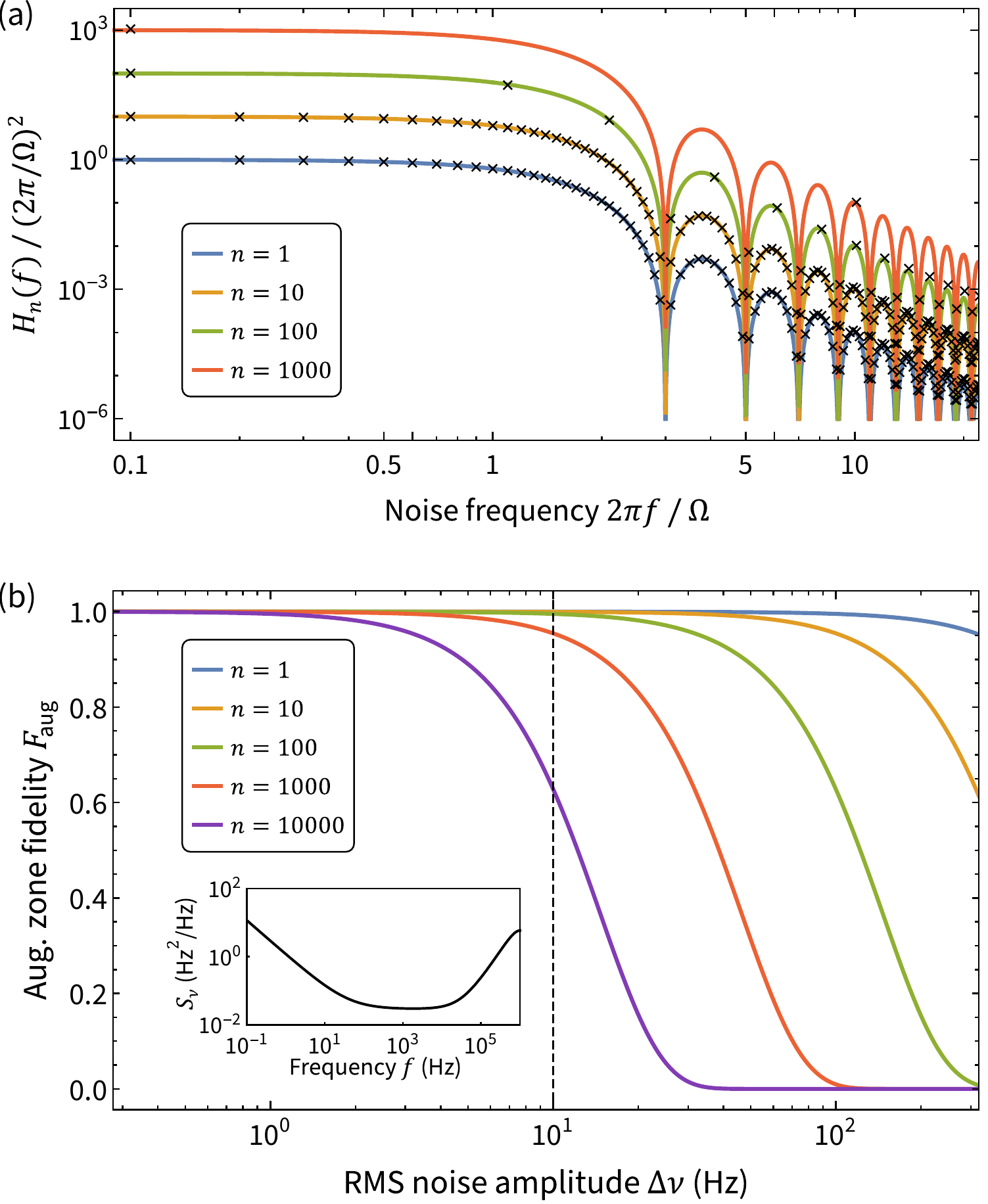}
    \caption{(a) Laser frequency noise transfer function $H_n(f)$ for an augmentation zone with $n$ pulses applied from alternating directions. The analytic expression agrees well with the data (black crosses), which are numerically calculated by solving the Schr\"odinger equations in the Hilbert space with all possible external momentum states. (b) Augmentation zone fidelity $F_{\textup{aug}}$ as a function of the rms laser frequency noise $\Delta\nu$ for various sequence lengths $n$, at a Rabi frequency of $\Omega=2\pi\times1~\text{kHz}$. Inset: Frequency noise spectral density at $\Delta\nu=10~\text{Hz}$ (indicated by dashed line). We vary the overall amplitude of this spectrum to set the rms noise amplitude.}
    \label{fig:transfer-function-and-fidelity}
\end{figure}

For this analysis, we vary the strength of the noise model by an overall scaling factor but keep its structure fixed, which is pink below $100~\text{Hz}$ and white above. We also include high-frequency noise above $10~\text{kHz}$ due to limitation of the servo gain, which is observed in cavity-stabilized laser systems~\cite{chiarotti2022practical,sterr2009ultrastable}. Note, however, that this servo-induced noise is outside the bandwidth of interest set by the Rabi frequency, and thus strongly suppressed by the roll-off of $H_n(f)$ when $2\pi f>\Omega$. We define the rms noise amplitude of the laser by integrating $S_\nu(f)$ from $f_l$ to $f_u$.

\begin{figure}[tb]
    \centering
    \includegraphics[width=\linewidth]{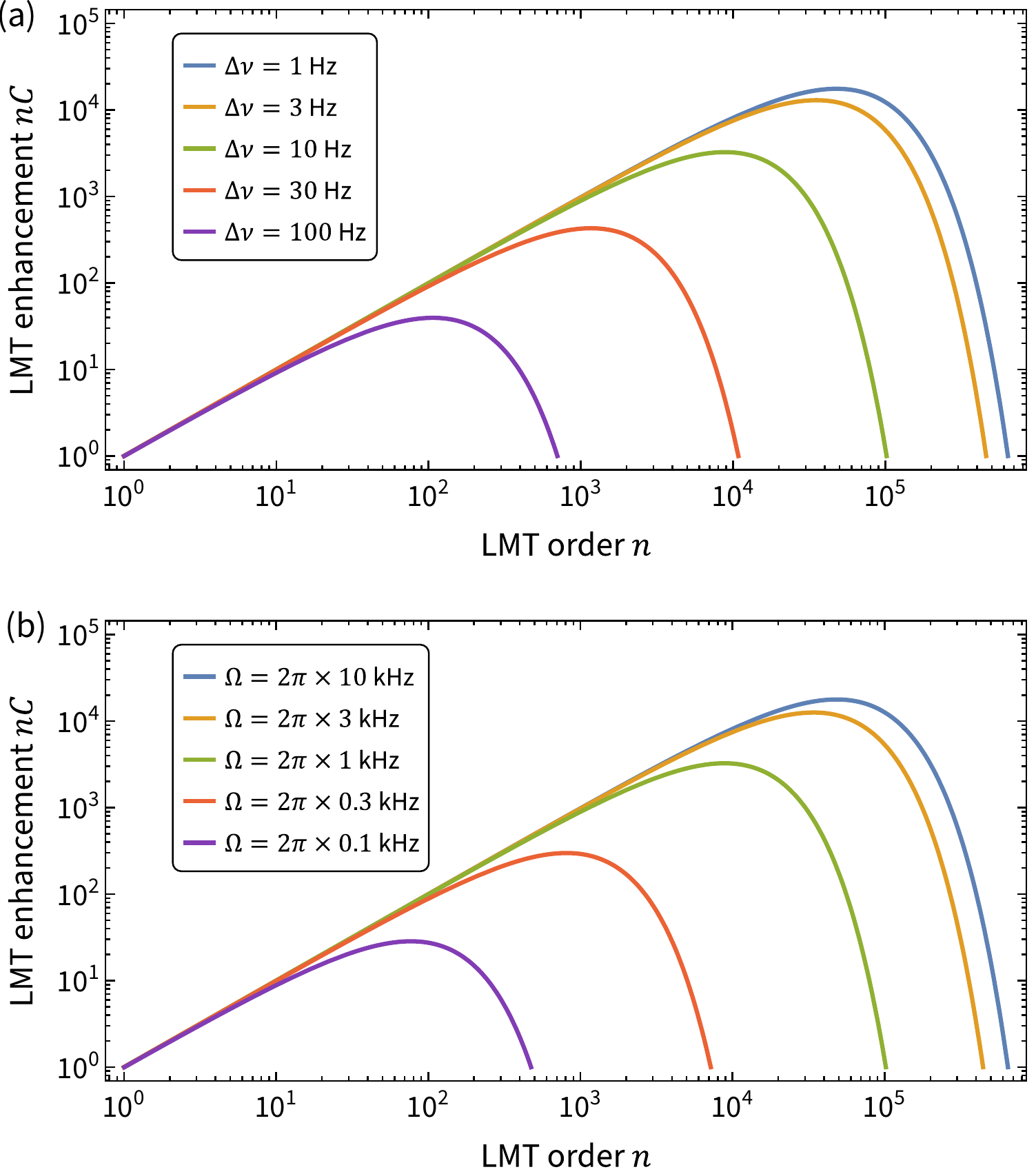}
    \caption{LMT enhancement of a clock atom interferometer in the presence of laser frequency noise as a function of LMT order $n$. Other loss mechanisms are assumed to introduce an additional pulse inefficiency of $10^{-5}$. (a) Comparison of different rms laser frequency noise  amplitudes $\Delta\nu$ for the same Rabi frequency $\Omega=2\pi\times1~\text{kHz}$. (b) Comparison of different Rabi frequencies using the same rms noise amplitude $\Delta\nu = 10~\text{Hz}$. In both panels, densely packed pulses and zero interrogation time are assumed.}
    \label{fig:enhancement-vs-n}
\end{figure}

Finally, we study a full $n\hbar k$ interferometer sequence, which is constructed by inserting four augmentation zones of length $n-1$ into a Mach-Zehnder interferometer. Besides laser frequency noise, we assume an additional pulse inefficiency of $10^{-5}$ from other loss mechanisms. The performance of the interferometer is evaluated by its LMT enhancement, defined as $n\,C$, which is proportional to the interferometer phase sensitivity. The existence of an optimal $n$ for a given $\Delta\nu$ and $\Omega$, as reported previously~\cite{chiarotti2022practical}, is observed in \figref{fig:enhancement-vs-n}. However, since population loss scales linearly rather than quadratically with respect to $n$, greater sensitivity enhancement is expected at a much larger optimal $n$. For a laser with $\Delta\nu=10~\text{Hz}$ and $\Omega=2\pi\times1~\text{kHz}$, a peak LMT enhancement $n\,C=3.3\times10^3$ is achieved at $n=8.8\times10^3$. This is consistent with the requirements for $1000\,\hbar k$ atom optics estimated for the MAGIS-100 detector~\cite{abe2021matter}. Moreover, higher Rabi frequencies and lower laser frequency noise can be achieved with current laser technology.

\section{Discussion}
\label{sec:discussion}
We find that for clock atom interferometers on narrowband transitions, such as \SrClock in strontium, the population loss in the main paths is proportional to the LMT order $n$. The interference background of the parasitic paths is bounded by a constant independent of $n$. The resulting loss of sequence fidelity grows linearly with $n$, and is a result of the momentum separation between the parasitic states and the target states when driven by $n$ pulses from alternating directions. This is in sharp contrast to the case of interrogating a two-level system $n$ times from the same direction, where the amplitude error in the wavefunction is subject to all subsequent pulses and coherently accumulates over time, leading to a fidelity loss that grows as $n^2$~\cite{green2013arbitrary}. Note that spontaneous emission loss also scales linearly with $n$ and, on such transitions, has a similar magnitude to the frequency-noise-induced loss for practical laser parameters $\Delta\nu=10~\text{Hz}$ and $\Omega=2\pi\times1~\text{kHz}$. This suggests that laser frequency noise does not pose a constraint on the contrast or phase sensitivity of narrowband LMT clock atom interferometers. Furthermore, frequency stabilization techniques that reduce the rms noise amplitude to below $\Delta\nu=1~\text{Hz}$ are readily available today.

Although this work is focused on the narrowband limit where the pulses are perfectly velocity selective, our results can be extended to broadband interferometers, where each pulse can address multiple Doppler classes. A finite Rabi frequency leads to off-resonant excitation, generating additional parasitic paths that are not included in this analysis. One possible mitigation is to use specifically tuned Rabi frequencies that prevent state changes of off-resonant parasitic paths with momentum $\pm 2\hbar k$ around the main paths, and strongly suppress state changes of paths that are farther Doppler detuned~\cite{rudolph2020large}. It is also possible to extend our analysis to clock atom interferometers where each $\pi$ pulse is intentionally made velocity insensitive to impart momentum to both the upper and the lower arms~\cite{rudolph2020large}. In this scenario, the argument for a linearly growing population loss is still valid for the main paths. However, more parasitic paths will be generated as the pulses are not velocity selective, leading to an increasing number of pulse error combinations for each $m$ term in \cref{eq:series-parasitic-expanded-upper}. Despite this, we note that if an error is made in an accelerating augmentation zone, the resulting parasitic wavefunction will be decelerated by subsequent pulses, and vice versa, due to the alternation of pulse direction. As a result, the majority of the parasitic paths lead to momentum states different from the main paths, and terminate far from the detection region. More importantly, as the LMT order $n$ increases, the broadband assumption becomes less valid, and the interferometer will enter the intermediate band~\cite{wilkason2022atom}. In this scenario, parasitic paths spawned many photon recoils away are sufficiently Doppler detuned to be treated by the formalism we have introduced here.

Beyond laser frequency noise, other sources of pulse inefficiency such as beam intensity fluctuations~\cite{legouet2008limits}, finite beam size~\cite{mielec2018atom}, and magnetic field errors~\cite{hu2017mapping} can induce transfer inefficiency at each pulse that decreases the population in the main paths and generates parasitic paths. The results presented here also apply to these sources of error, leading to the same conclusion about fidelity loss scaling. For narrowband LMT clock atom interferometers with these noise sources, we do not expect the interference background from parasitic paths to dominate over the amplitude loss in the main paths. Finally, many above-mentioned sources of pulse inefficiency, including laser frequency noise, can be alleviated by the implementation of composite pulse, optimal control, or pulse-shaping techniques tailored to one or several specific types of errors, at the expense of a longer pulse duration~\cite{dunning2014composite,zanon2022generalized,saywell2020optimal,correia2022quantum,chen2023enhancing,fang2018improving,saywell2018optimal,butts2013efficient,levitt2007composite,levitt1982symmetrical,berg2015composite}. Such techniques have the potential to suppress both main path losses and parasitic path amplitudes, thereby further relaxing the laser stability requirement beyond the results in this paper.

\section*{Acknowledgments}
This work was supported in part by the National Science Foundation QLCI Award No. OMA-2016244, the Gordon and Betty Moore Foundation Grant No. GBMF7945, and by the National Science Foundation under Award No. 2409710.

\appendix

\section{\MakeUppercase{Combinatorial analysis of parasitic paths}}
\label{appendix:combinatorial}
In this Appendix, we analyze the structure of parasitic paths in a narrowband clock atom interferometer and give a combinatorial treatment to the total number of parasitic paths $N_{p,x}(m,n)$ that are relevant to the interference background.

For a generic state $\ket{\psi}=\ket{s,p}$ in the interferometer, to facilitate the analysis, we assign values to its internal level $s$ as $s=1$ (ground) and $s=-1$ (excited). We denote a $\pi$ pulse as a dimensionless tuple $(d,f)$ and a $\pi/2$ pulse as $(d,f)_{\frac{\pi}{2}}$. In this notation, $d=\pm1$ represents the direction of the pulse, and $f=\tilde{f}d$, where $\tilde{f}$ is the laser detuning in the free-falling frame in units of $\hbar k^2/m$. The state $\ket{s,p}$ is coupled to the state $\ket{s',p'}$ where $s'=-s$ and $p'=p+sd$, if and only if the laser detuning compensates for the Doppler shift of the moving atoms. This resonance condition can be written as
\begin{equation}
    f=\frac{p+p'}{2}=p+\frac{1}{2}sd,
\end{equation}
and the wavefunction after interacting with an imperfect $\pi$ pulse is $\ket{\psi'}=e^{i\phi}\sqrt{1-\delta\!P}\ket{s',p'}+\alpha\ket{s,p}$, where $\alpha$ is a small complex amplitude, $\delta\!P=|\alpha|^2$ is the pulse inefficiency, and $\phi$ is the phase shift imprinted due to the atom-light interaction.

We denote a pulse sequence as a linked list, with arrows between its constituent pulses or subsequences to indicate time ordering. A Mach-Zehnder interferometer, for example, is expressed as
\begin{equation}
    L_{\text{MZ}}=\left(1,\tfrac{1}{2}\right)_{\frac{\pi}{2}}\stackrel{T}{\rightarrow}\left(1,\tfrac{1}{2}\right)\stackrel{T}{\rightarrow}\left(1,\tfrac{1}{2}\right)_{\frac{\pi}{2}},
\end{equation}
where $T$ represents a finite interrogation time. The narrowband $n\hbar k$ interferometer shown in Fig.~\ref{fig:interferometer-diagrams} is structured by inserting the augmentation sequence $L_{\text{aug}}=(-1,\frac{3}{2})\rightarrow(1,\frac{5}{2})\rightarrow\cdots\rightarrow\left((-1)^{n-1},n-\frac{1}{2}\right)$ and its reversed sequence, denoted as $L_{\text{aug}}^{-1}$, into a Mach-Zehnder interferometer:
\begin{align}
    L=&\left(1,\tfrac{1}{2}\right)_{\frac{\pi}{2}}\rightarrow L_{\text{aug}}\stackrel{T}{\rightarrow}L_{\text{aug}}^{-1}\rightarrow\left(1,\tfrac{1}{2}\right)\notag\\
    &\rightarrow L_{\text{aug}}\stackrel{T}{\rightarrow}L_{\text{aug}}^{-1}\rightarrow\left(1,\tfrac{1}{2}\right)_{\frac{\pi}{2}}.
    \label{eq:appendix-a.lmt-sequence}
\end{align}
For convenience, we assume perfect beamsplitters and focus on the $\pi$ pulse section of $L$, which contains $4n{-}3$ pulses whose directions and detunings are indicated by the arrows in \figref[a]{fig:parasitic-paths}, with the following analytic forms
\begin{align}
    d_j&=\left\{\begin{array}{ll}(-1)^j&\quad 1\leqslant j<n\\
    (-1)^{2n-j+1}&\quad n\leqslant j<2n-1\\
    1&\quad j=2n-1\\
    (-1)^{j-2n-1}&\quad 2n-1<j\leqslant 3n-2\\
    (-1)^{4n-j}&\quad 3n-2<j\leqslant 4n-3
    \end{array}\right.\label{eq:appendix-a.pulse-direction}\\
    f_j&=\left\{\begin{array}{ll}j+\frac{1}{2}&\,\,\quad 1\leqslant j<n\\
    2n-j-\frac{1}{2}&\,\,\quad n\leqslant j<2n-1\\
    \frac{1}{2}&\,\,\quad j=2n-1\\
    j-2n+\frac{3}{2}&\,\,\quad 2n-1<j\leqslant 3n-2\\
    4n-j-\frac{3}{2}&\,\,\quad 3n-2<j\leqslant 4n-3
    \end{array}\right.\label{eq:appendix-a.pulse-detuning}
\end{align}
where we show the four augmentation zones as piecewise functions, along with the center mirror pulse. Note that each augmentation zone has monotonic frequency with respect to the pulse index $j$.

Here, we use the examples in \figref[c]{fig:parasitic-paths} to show how multiple pulse errors can generate a parasitic path with the same final momentum as one of the main paths in order to contribute to the interference background. In the examples, a pulse error occurs at $j=2$, and the error state becomes off resonant with subsequent pulses in the same augmentation zone. The resulting parasitic path drifts at a constant momentum until pulse $j'=5$ occurs in the next augmentation zone. This pulse is on resonance since $d_5=d_2$ and $f_5=f_2$. From this point, the parasitic path will either undergo a transition or not. If an additional error occurs here and it fails to make the transition at $j'=5$, then the rest of the parasitic path overlaps with the main path, either to the end of the interferometer or until another pulse error is made. The former results in the $m=2$ path in \figref[c]{fig:parasitic-paths}, and examples of the latter are given by $m=4,6$ in \figref[c]{fig:parasitic-paths}. On the other hand, if the transition at $j'=5$ does occur, then the parasitic path is once again off resonance with subsequent pulses until another resonant pulse $j''=9$ in the next augmentation zone, and the process repeats.

Note that when $f_j=f_{j'}$, \cref{eq:appendix-a.pulse-direction,eq:appendix-a.pulse-detuning} guarantee that $d_j=d_{j'}$. We thus conclude that a parasitic path terminates at the same momentum as one of the main paths if two pulse errors occur with the same detuning. As shown in \figref[c]{fig:parasitic-paths}, this can be generalized to an even number of pulse errors arranged in equal-detuning pairs. In addition, the center mirror pulse at $j_{\text{c}}=2n-1$ does not have an equal-detuning counterpart, but an error is allowed to occur on this pulse, with examples shown in \figref[b]{fig:parasitic-paths}.  Consequently, if a parasitic path fulfilling the momentum constraint contains $m$ pulse errors, they have to be grouped into $\floor{m/2}$ pairs. Since there are $\mathcal{O}(n)$ ways to arrange each pair in an LMT sequence of $4n-3$ pulses, the total number of parasitic paths is bounded by $N_p(m,n)=\mathcal{O}(n^{\floor{m/2}})$. Finally, the narrowband interferometer structure in \cref{eq:appendix-a.lmt-sequence} contains four augmentation zones, which limits the pulse errors to be no more than three pairs, or equivalently, $m\leqslant6$.

To account for the final position of the parasitic paths, we make the worst-case assumption of zero interrogation time $T$, and consider a path that terminates with the correct momentum but not necessarily the correct position. Note that $m=1$ is only possible when the upper arm encounters an error at the center mirror pulse, resulting in a parasitic path that is nominally on resonance with all subsequent pulses, even though they are frequency tuned to the lower arm. As shown in \figref[b]{fig:parasitic-paths}, this parasitic path experiences twice the displacement as the main paths in the free-falling frame, and does not terminate at the detection region for large enough $n$.

We thus focus on paths with $m>1$ where at least one pair of pulse errors occurs. We denote the pulse indices in a pair as $j$ and $\bar{j}$. Note that up to the choice of which augmentation zone, $\bar{j}$ is completely determined by $j$ based on the $f_j=f_{\bar{j}}$ constraint. Then we can express the relative position of a parasitic path with respect to the main path as a function of the integer-valued free variables $j_1,\cdots,j_{\floor{m/2}}$, the first indices of the $\floor{m/2}$ pairs of errors
\begin{equation}
    x_{\text{para}}-x_{\text{main}}=\delta x\left(j_1,\cdots,j_{\floor{m/2}}\right),
\end{equation}
which is also true for odd $m$ since the center mirror pulse $j_\text{c}$ is not a free variable. Geometrically, this represents the signed area of the shaded regions in the momentum-time diagrams in \figref{fig:parasitic-paths}, which must be a quadratic function of the free variables. Moreover, the relative position $\delta x$ is proportional to the recoil velocity $v_r$ and the pulse spacing $\delta t$. We thus define the quadratic form
\begin{equation}
    Q\left(j_1,\cdots,j_{\floor{m/2}}\right)\equiv\frac{1}{v_r\,\delta t}\delta x\left(j_1,\cdots,j_{\floor{m/2}}\right)
\end{equation}
as the dimensionless relative position. A parasitic path will interfere with the main paths if
\begin{equation}
    \left|Q\left(j_1,\cdots,j_{\floor{m/2}}\right)\right|\leqslant\frac{w}{v_r\,\delta t},
    \label{eq:appendix-a.position-constraint}
\end{equation}
where $w$ is the radius of the detection region. We assume $w=1~\text{mm}$, so the wavepacket spread is small in comparison and the paths can be treated classically. Inequality~\ref{eq:appendix-a.position-constraint} poses further constraints on the $\floor{m/2}$ free parameters, leading to a reduction of feasible parasitic paths from $N_p(m,n)$ to $N_{p,x}(m,n)$.

To deduce $N_{p,x}(m,n)$, we calculate the available choices of $j_1$ that satisfy the position constraint, while keeping the other indices free. Assuming some $j_1^*\in\mathbb{R}$ exists that satisfies $Q(j_1^*,\cdots,j_{\floor{m/2}})=0$, we expand $Q$ about $j_1^*$ and obtain
\begin{equation}
    Q\left(j_1,\cdots,j_{\floor{m/2}}\right)=B~\delta j_1+A~\delta j_1^2,\label{eq:appendix-a.quadratic-expansion}
\end{equation}
where $\delta j_1=j_1-j_1^*$, and we have defined
\begin{equation}
    A\equiv\left.\frac{\partial^2Q}{\partial j_1^2}\right|_{j_1=j_1^*}, \quad
    B\equiv\left.\frac{\partial Q}{\partial j_1}\right|_{j_1=j_1^*}.
\end{equation}
Note that the expansion is exact since $Q$ is quadratic with respect to $j_1$. In fact, depending on which augmentation zones $j_1$ and $\bar{j}_1$ belong to, we can write down explicitly how $Q$ depends on $j_1$ by calculating the area of the shaded regions in \figref{fig:parasitic-paths}. We summarize the results in \cref{table:appendix-a.quadratic-expansion}, and note that $|A|=0$ or $|A|=2$ in all scenarios.
\begin{table}[t]
    \caption{Dependence of the dimensionless relative position $Q$ on the first error index $j_1$. Augmentation zones 1,2 refer to the $L_{\text{aug}}$ and $L^{-1}_{\text{aug}}$ before the center mirror pulse in \cref{eq:appendix-a.lmt-sequence}, and 3,4 refer to those after the center mirror pulse. The $j_1$-independent terms are omitted.}
    \label{table:appendix-a.quadratic-expansion}
    \begin{tabular}{M{0.17\linewidth}M{0.17\linewidth}M{0.35\linewidth}M{0.1\linewidth}M{0.13\linewidth}}
    \hline
    \vspace{2pt} Zone of $j_1$ & Zone of $\bar{j}_1$ & $Q$ & $A$ & $B$ \\[2pt]
    \hline
    \vspace{2pt} 1 & 2 & $-j_1^2+2nj_1+\cdots$ & $-2$ & $2n$ \\[2pt]
    \vspace{2pt} 1 & 3 & $2nj_1+\cdots$ & $0$ & $2n$ \\[2pt]
    \vspace{2pt} 1 & 4 & $-j_1^2+4nj_1+\cdots$ & $-2$ & $4n$\\[2pt]
    \vspace{2pt} 2 & 3 & $j_1^2-4nj_1+\cdots$ & $2$ & $-4n$ \\[2pt]
    \vspace{2pt} 2 & 4 & $-2nj_1+\cdots$ & $0$ & $-2n$ \\[2pt]
    \vspace{2pt} 3 & 4 & $-j_1^2+6nj_1+\cdots$ & $-2$ & $6n$ \\[2pt]
    \hline
    \end{tabular}
\end{table}

Inequality~(\ref{eq:appendix-a.position-constraint}) can be rewritten in terms of $\delta j_1$ as
\begin{equation}
    \left|A~\delta j_1^2+B~\delta j_1\right|\leqslant W,
    \label{eq:appendix-a.position-constraint-simplified}
\end{equation}
where $W\equiv\frac{w}{v_r\,\delta t}$. We first consider the $|A|=2$ scenario, where we define $\Delta j_1\equiv\frac{|A|}{W}(\delta j_1+\frac{B}{2A})$ and $X\equiv\frac{B^2}{4|A|W}$. When $X\leqslant1$, the inequality simplifies to
\begin{equation}
    \Delta j_1\in\left[-\sqrt{X+1},\sqrt{X+1}\right]
\end{equation}
and when $X>1$,
\begin{equation}
    \Delta j_1\in\left[-\sqrt{X+1},-\sqrt{X-1}\right]\bigcup\left[\sqrt{X-1},\sqrt{X+1}\right].
\end{equation}
In either case, the total length of the interval does not exceed $2\sqrt{2}$. We thus conclude that $\delta j_1$ is confined within an interval of length
\begin{equation}
    D\leqslant2\sqrt{2}\,\frac{W}{|A|}\leqslant\frac{\sqrt{2}\,w}{v_r\,\delta t}.
\end{equation}

In the $|A|=0$ scenario, notice from \cref{table:appendix-a.quadratic-expansion} that $|B|=2n$; then, Inequality~(\ref{eq:appendix-a.position-constraint-simplified}) reduces to
\begin{equation}
    |\delta j_1|\leqslant\frac{W}{2n}
\end{equation}
and the interval length still satisfies $D\leqslant\frac{\sqrt{2}\,w}{v_r\,\delta t}$ for $n\geqslant1$.

With the position constraint imposed, $j_1$ can only take integer values in an interval of length $D$ around $j_1^*$, as opposed to $\mathcal{O}(n)$ choices in the interferometer sequence. As a result, although $N_p(m,n)$ parasitic paths terminate with the correct momentum, only $N_{p,x}(m,n)=\frac{D}{n}N_p(m,n)=\frac{w}{v_r\,\delta t}\times\mathcal{O}(n^{\floor{m/2}-1})$ of them can enter the detection region and affect the interferometer contrast. Note that the choice of $j_1$ being the constrained index is arbitrary, and is equivalent to constraining any index among $j_1,\cdots,j_{\floor{m/2}}$.

We incorporate this form of $N_{p,x}(m,n)$ into the weighted sum of wavefunction amplitudes in \cref{eq:total-parasitic-wavefunction} assuming constructive interference among all relevant paths. The total contribution from parasitic paths to the relevant states $\ket{\text{g},0}$ and $\ket{\text{e},1}$ is then given by \cref{eq:total-parasitic-wavefunction-bound} in the paper.

\section{\MakeUppercase{Analytic derivation of frequency noise transfer function}}
\label{appendix:analytic}
Phase sensitivity of an atom interferometer in the presence of laser frequency noise has been well studied~\cite{cheinet2008measurement}. Recently, there have been works based on the filter function formalism and perturbative solution of the master equation that analyze qubit gate fidelity under a noisy laser~\cite{day2022limits,jiang2023sensitivity}. In this Appendix, we analytically derive the amplitude response of the interferometer wavefunction when addressed by a $\pi$ pulse with frequency noise. We take a different approach of perturbatively expanding the time evolution operator, and obtain the single-pulse frequency noise transfer function. We then put our results into the context of a narrowband clock atom interferometer and extend the analysis to a sequence of $n$ pulses from alternating directions.

We consider a single $\pi$ pulse with time-dependent coupling $\Omega(t)$ driving the transition between two states $\ket{1}$ and $\ket{2}$ split by an energy difference of $\hbar\omega_0$. The light field is assumed to have nominal angular frequency $\omega$ and a small frequency noise term $\delta\omega(t)$, where $\delta\omega(t)$ is an arbitrary function. The Hamiltonian after the rotating wave approximation can be written as
\begin{equation}
    \hat{H}=\frac{1}{2}\hbar\Omega(t)\,e^{i\left((\omega-\omega_0)t+\varphi(t)\right)}\ket{1}\!\bra{2}+\text{H.c.},
\end{equation}
where $\varphi(t)=\int_{t_0}^t\delta\omega(t')\,dt'+\varphi_0$ and $t_0$ marks the beginning of the atom-light interaction. We begin by transforming the Hamiltonian into a rotating frame using the unitary transformation $\hat{R}(t)=e^{-\frac{i}{2}\hat{\sigma}_z\varphi(t)}$, where $\hat{\sigma}_x,\hat{\sigma}_y,\hat{\sigma}_z$ are the Pauli matrices. Assuming the laser is nominally on resonance, the transformed Hamiltonian can be expressed as
\begin{equation}
    \hat{H}_R=\hat{R}\hat{H}\hat{R}^\dag+i\hbar\,\partial_t\hat{R}\,\hat{R}^\dag=\hat{H}_0+\hat{\delta\!H},
\end{equation}
where $\hat{H}_0=\frac{1}{2}\hbar\Omega(t)\hat{\sigma}_x$ is the noise-free Hamiltonian, and $\hat{\delta\!H}=\frac{1}{2}\hbar\,\delta\omega(t)\hat{\sigma}_z$ is treated as a perturbation. The interaction picture Hamiltonian is obtained using the time evolution operator $\hat{U}_0(t,t_0)=e^{-\frac{i}{\hbar}\int_{t_0}^t\hat{H}_0(t')\,dt'}$ as
\begin{equation}
    \hat{H}_{\text{int}}=\hat{U}_0^\dag\,\hat{\delta\!H}\,\hat{U}_0=\frac{1}{2}\hbar\,\delta\omega(t)\left(\hat{\sigma}_y\sin A(t)+\hat{\sigma}_z\cos A(t)\right),
\end{equation}
where $A(t)=\int_{t_0}^t\Omega(t')\,dt'$ is the pulse area. For a $\pi$ pulse with duration $\tau$, $A(t_0+\tau)=\pi$. We expand the interaction picture time evolution operator with a Dyson series truncated at second order in the perturbation and obtain
\begin{align}
    \hat{U}_{\text{int}}\approx&~1-\frac{i}{\hbar}\int_{t_0}^t\hat{H}_{\text{int}}(t')\,dt'\notag\\
    &+\left(-\frac{i}{\hbar}\right)^2\int_{t_0}^t\int_{t_0}^{t'}\hat{H}_{\text{int}}(t')\,\hat{H}_{\text{int}}(t'')\,dt''dt',
\end{align}
where the approximation is valid as long as $|\delta\omega(t)|\ll\Omega(t)$. The state of the atom at any given time $t$ is $\ket{\psi(t)}=\hat{U}(t,t_0)\ket{\psi(t_0)}$, where $\hat{U}(t,t_0)=\hat{R}^\dag(t)\hat{U}_0\hat{U}_\text{int}\hat{R}(t_0)$ is the complete time evolution operator. Assuming the initial state to be $\ket{1}$, we can write the amplitude in the target state $\ket{2}$ after the $\pi$ pulse of duration $\tau$ as
\begin{equation}
    \braket{2|\psi(t_0+\tau)}=-ie^{-i\bar{\varphi}}(1-iu_1-u_2),
\end{equation}
where we have defined
\begin{align}
    \bar{\varphi}&\equiv\frac{1}{2}\left(\varphi(t_0+\tau)+\varphi(t_0)\right),\\
    u_1&\equiv\frac{1}{2}\int_{t_0}^{t_0+\tau}\cos A(t')\,\delta\omega(t')\,dt',\\
    u_2&\equiv\frac{1}{4}\int_{t_0}^{t_0+\tau}\!\!\!\int_{t_0}^{t'}\cos(A(t')-A(t''))\,\delta\omega(t')\,\delta\omega(t'')\,dt''dt'.
\end{align}
The population transfer error up to second order in $\delta\omega(t)$ can be written as
\begin{equation}
    \delta\!P=1-\left|\braket{2|\psi(t_0+\tau)}\right|^2\approx 2u_2-u_1^2.
\end{equation}
After plugging in the expressions for $u_1$ and $u_2$,
\begin{equation}
    \delta\!P=\!\left(\frac{1}{2}\!\int_{t_0}^{t_0+\tau}\!\!\!\!\!\sin A(t)\,\delta\omega(t)\,dt\right)^2\!\!=\!\left(\int_{-\infty}^{+\infty}\!\!\!\!h_1(t)\,\delta\nu(t)\,dt\right)^2,
    \label{eq:appendix-b.population-loss-time-domain}
\end{equation}
where $\delta\nu(t)=\frac{\delta\omega(t)}{2\pi}$. The second-order time response function is defined as
\begin{equation}
    h_1(t)\equiv\pi\sin\!\left(\int_{t_0}^{t}\!\Omega(t')\,dt'\!\right)\!\left(\Theta(t-t_0)-\Theta(t-(t_0+\tau))\right),
\end{equation}
with $\Theta(t)$ being the unit step function. Using Fourier transforms $\widehat{h_1}(f)=\int_{-\infty}^{+\infty}h_1(t)\,e^{-i\cdot2\pi ft}\,dt$ and $\widehat{\delta\nu}(f)=\int_{-\infty}^{+\infty}\delta\nu(t)\,e^{-i\cdot2\pi ft}\,dt$, we write the population error in \cref{eq:appendix-b.population-loss-time-domain} as
\begin{equation}
    \delta\!P=\int_{-\infty}^{+\infty}\!\!\!\int_{-\infty}^{+\infty}\widehat{h_1}(f)\,\widehat{h_1}^*\!(f')\,\widehat{\delta\nu}^*\!(f)\,\widehat{\delta\nu}(f')\,df'df.
    \label{eq:appendix-b.population-loss-frequency-domain}
\end{equation}
The frequency response function $\widehat{h_1}(f)$ can be analytically calculated assuming a $\pi$ pulse with $\Omega(t)=\Omega=\pi/\tau$:
\begin{equation}
    \widehat{h_1}(f)=\pi\Omega\left(\frac{1+e^{-i\cdot2\pi f\frac{\pi}{\Omega}}}{\Omega^2-(2\pi f)^2}\right)e^{-i\cdot2\pi ft_0}.
    \label{eq:appendix-b.frequency-response-function}
\end{equation}

To obtain the transfer function with respect to the frequency noise power spectral density, we assume $\delta\nu(t)$ to be a stationary process and calculate the ensemble average of $\widehat{\delta\nu}^*\!(f)\,\widehat{\delta\nu}(f')$,
\begin{align}
    &\expect{\widehat{\delta\nu}^*\!(f)\,\widehat{\delta\nu}(f')}\notag\\
    =&\int_{-\infty}^{+\infty}e^{i\cdot2\pi(f-f')t}\,dt\int_{-\infty}^{+\infty}\expect{\delta\nu^*(t-\tau)\,\delta\nu(t)}\,e^{-i\cdot2\pi f\tau}\,d\tau\notag\\
    =&~\tilde{S}_{\nu}(f)\,\delta(f-f'),
    \label{eq:appendix-b.ensemble-average}
\end{align}
where the last step uses the Wiener-Khinchin theorem and identifies the Fourier transform of the autocorrelation function $\rho(\tau)=\expect{\delta\nu^*(t-\tau)\,\delta\nu(t)}$ as the two-sided power spectral density $\tilde{S}_{\nu}(f)$. We apply \cref{eq:appendix-b.ensemble-average} to the ensemble average of \cref{eq:appendix-b.population-loss-frequency-domain} and obtain
\begin{equation}
    \expect{\delta\!P}=\int_{0}^{\infty}\left|\widehat{h_1}(f)\right|^2S_{\nu}(f)\,df,\label{eq:appendix-b.ensemble-averaged-loss}
\end{equation}
where we have switched to the one-sided power spectral density $S_{\nu}(f)=2\tilde{S}_{\nu}(f)$. We recognize $|\widehat{h_1}(f)|^2$ as the single-pulse frequency noise transfer function $H_1(f)$, which we write analytically as \cref{eq:transfer-function} in the paper using \cref{eq:appendix-b.frequency-response-function}. Since parasitic paths are negligible, due to the linear scaling of population error with respect to the number of pulses, the transfer function for $n$ sequential $\pi$ pulses applied from alternating directions is $H_n(f)=nH_1(f)$:
\begin{equation}
    H_n(f)=n\left(\frac{2\pi}{\Omega}\right)^2\left(\frac{\Omega^2\cos\left(\pi f\,\frac{\pi}{\Omega}\right)}{\Omega^2-(2\pi f)^2}\right)^2.
    \label{eq:appendix-b.transfer-function-n}
\end{equation}

Note that in the case where the noise is concentrated below the Rabi frequency, $H_n(f)$ is approximately a constant in the noise bandwidth, and we have
\begin{equation}
    \expect{\delta\!P} \approx H_n(0)\int_0^{\infty}S_{\nu}(f)\,df=n\frac{\Delta\nu^2}{(\Omega/2\pi)^2}
\end{equation}
In practice, this integral is performed over a finite bandwidth between $f_l$ and $f_u$ introduced in Sec.\,\ref{sec:transfer-function}, and $\Delta\nu=\sqrt{\int_{f_l}^{f_u}S_{\nu}(f)\,df}$. This result is consistent with the analysis in Ref.~\cite{abe2021matter}.

\bibliographystyle{apsrev4-2-titles}
\bibliography{bibliography}
\end{document}